\documentclass{article}

\usepackage{arxiv}

\usepackage{mathptmx}
\usepackage{amsmath}
\usepackage{amssymb}
\usepackage{amsthm}

\usepackage[utf8]{inputenc} 
\usepackage[T1]{fontenc}    
\usepackage{hyperref}       
\usepackage{url}            
\usepackage{booktabs}       
\usepackage{amsfonts}       
\usepackage{nicefrac}       
\usepackage{microtype}      
\usepackage{lipsum}

\usepackage{optidef}
\usepackage{algorithm}
\usepackage{algpseudocode}
\usepackage{algorithmicx}
\usepackage{enumitem}
\usepackage{tikz} 

\usepackage{caption}
\usepackage{subcaption}
\usepackage{centernot}
\usepackage{xcolor}

\newtheorem{Theorem}{Theorem}

\newtheorem{Proposition}{Proposition}
\newtheorem{Definition}{Definition}

\newcommand\independent{\protect\mathpalette{\protect\independenT}{\perp}}
\def\independenT#1#2{\mathrel{\rlap{$#1#2$}\mkern2mu{#1#2}}}
\newcommand{\notindependent}{\centernot{\independent}}

\DeclareMathAlphabet{\mathcal}{OMS}{cmsy}{m}{n}
\newcommand{\argmin}{\mathop{\mathrm{argmin}}}

\title{On efficient covariate adjustment selection in causal effect estimation}

\author{
 Hongyi Chen \\
  Tilburg University \\
  the Netherlands \\
   \And
 Maurits Kaptein\\
  Tilburg University\\
  the Netherlands \\
}

\begin{document}
\maketitle
\begin{abstract}
In order to achieve unbiased and efficient estimators of causal effects from observational data, covariate selection for confounding adjustment becomes an important task in causal inference. Despite recent advancements in graphical criterion for constructing valid and efficient adjustment sets, these methods often rely on assumptions that may not hold in practice. We examine the properties of existing graph-free covariate selection methods with respect to both validity and efficiency, highlighting the potential dangers of producing invalid adjustment sets when hidden variables are present. To address this issue, we propose a novel graph-free method, referred to as CMIO, adapted from Mixed Integer Optimization (MIO) with a set of causal constraints. Our results demonstrate that CMIO outperforms existing state-of-the-art methods and provides theoretically sound outputs. Furthermore, we present a revised version of CMIO capable of handling the scenario in the absence of causal sufficiency and graphical information, offering efficient and valid covariate adjustments for causal inference.
\end{abstract}


\section{Introduction}
Covariate selection holds great importance in various popular methodologies, including regression analysis, propensity score matching, and doubly robust methods, which aim to determine outcomes or causal effects resulting from different exposures, treatments, or interventions based on observational data. A covariate set that mediates all confounding factors affecting both exposure and outcome is deemed as a valid adjustment set. (A rigorous definition in mathematical terms will be provided later in the paper.) With a focus on sufficiency, research in recent decades has sought to examine the characteristics of valid adjustment sets that can reduce the variance of unbiased estimators, one example of such is a recent paper \cite{guo2022efficient} on efficient estimators for recursive linear structural equation models . Two key pieces of guidance have been established for comparing and defining efficient valid adjustment sets, dependent on the presence of causal graphs. The graphical criteria for evaluating the asymptotic variances of valid adjustment sets were first introduced in linear models by \cite{henckel2019graphical} and further expanded upon in more general settings of non-parametric models by \cite{rotnitzky2019efficient}. While non-graphical settings can only provide heuristic suggestions, such as including predictors of outcome variables \cite{witte2019covariate}. One of the significant benefits of the graphical criteria proposed is the ability to identify an optimal adjustment set that results in the unbiased causal effect estimator with the smallest asymptotic variance among all valid adjustment sets.

As significant as determining the composition of an efficient valid adjustment set is, so too is the challenge of selecting such a set from a vast array of covariates, particularly in the absence of a causal graph. Currently, many techniques for efficient covariate selection, which include nearly all algorithms underpinned by optimization methods, e.g. \cite{wang2012bayesian,shortreed2017outcome,talbot2015bayesian}, have overlooked the potential risk of invalidity that arises from mistakenly identifying the target set that is both valid and more efficient.These methods have been demonstrated to be effective under the assumption of causal sufficiency: no hidden variables allowed. Our research, however, highlights that these methods may become unreliable when applied to more realistic scenarios featuring hidden variables. As a result, it is crucial to assess the limitations of state-of-the-art methods not only in terms of their efficiency but also in terms of their validity in the presence of hidden variables. Additionally, the development of novel, robust methods and algorithms for efficient covariate adjustment selection that can handle challenging, real-world scenarios such as the existence of hidden variables is of immense value.

Motivated by the aforementioned observations, we present three contributions on two dimensions that are validity and efficiency of covariate selection methods of interests as followings.  
\begin{itemize}

    \item The first objective of this study is to delve into the properties of existing prediction-based covariate selection methods. Particular attention is given to the potential dangers of producing invalid adjustment sets when hidden variables are present and there is a lack of graphical information. (Section 3)

    \item Assuming causal sufficiency, although the existing methods have demonstrated accuracy, we propose a novel graph-free method which enhances the efficiency of these methods. The novel approach along with an instantiated algorithm, denoted as CMIO, is adapted from the Mixed Integer Optimization (MIO) with a set of causal constraints. In Section 4, we demonstrate that the output of CMIO is theoretically sound and will coincide with the optimal adjustment set with probability 1, provided that the sample size is sufficiently large and complies with a linear causal model. In Section 6, we conduct a comparison between the performance of CMIO and other state-of-the-art variable selection approaches for causal effect estimation using simulated data and show that CMIO outperforms the other methods. 
    
    \item Lastly, in Section 5, we revisit the scenario in the absence of causal sufficiency and graphical information, offering an algorithm for generating valid and efficient sets of covariates. Drawing upon the insights gained from our earlier work on CMIO, we present a revised version capable of handling such demanding circumstances and offer both efficient and valid sets of covariate adjustments for causal inference. The statistical properties of this revised method are rigorously demonstrated in the appendices. p
    
\end{itemize}

\section{Preliminaries}
Before presenting our primary results, we commence with a preliminary discussion of the underlying causal framework, the Structural Causal Model (SCM) as introduced by Pearl \cite{pearl2009causality}. Furthermore, we provide an explanation of the graph-based terminologies and definitions used to represent random variables, conditional dependencies, and direct causal effects. Additionally, we include a brief overview of relevant literature regarding efficient covariate selection methods for causal inference in this section.

\subsection{Graph terminology}
A graph is an ordered pair defined as $\mathcal{G}=(V,E)$ consisting of a vertices set V and an edges set E. In the context of graphical modelling, vertices represent random variables and edges encode probabilistic and causal relations between vertices associated with them. For convenience, we use terms vertices and variables interchangeably.

A \textit{(un)directed} graph is a type of graph which contains only (un)directed edges. Otherwise, it is called a \textit{mixed} or \textit{partially directed} graph. In particular, a directed graph absent of \textit{directed cycles} is known as a \textit{Directed Acyclic Graph} (DAG) that could be transformed into an undirected graph by removing all edge directions, referred as \textit{skeleton} of the DAG. Two vertices are \textit{adjacent} if they are linked with an edge. A \textit{directed path} is a sequence of distinct vertices which are successively adjacent by edges of the same direction. If there is a directed path from vertex $X$ to $Y$, then $X$ is an ancestor of $Y$ while $Y$ is a descendant of $X$. Moreover, if such a directed path is an edge, we call $X$ a parent of $Y$ and $Y$ a child of $X$. The sets of parents, children, ancestors and descendants of a vertex $X$ in graph $\mathcal{G} $ are denoted as $\mathbf{PA}(\mathcal{G},X),\mathbf{CH}(\mathcal{G},X),\mathbf{AN}(\mathcal{G},X)$ and $\mathbf{DE}(\mathcal{G},X)$ accordingly.

A graphical criterion designated as \textit{d-seperation} \cite{lauritzen1996graphical, pearl2009causality} specifies conditional independence relationships of a DAG comprehensively. If joint distribution of $\mathbf{X}$, $\mathbb{P}(\mathbf{X})$, contains all the conditional independence relationships encoded by a DAG $\mathcal{G}$, the distribution is said to be Markovian to $\mathcal{G}$. On the other hand, a distribution $\mathbb{P}(\mathbf{X})$ is said to be faithful to a graph $\mathcal{G}$ if every conditional independence relation of $\mathbb{P}(\mathbf{X})$ is entailed by that in $\mathcal{G}$ \cite{spirtes2000causation}. If a distribution is both Markovian and faithful with respect to a DAG, we call the DAG a perfect map of the distribution.

\subsection{Structural Causal Models}
A structural equation model (SEM) determines the marginal distribution of each variable in the random vector $\mathbf{X}$=$(X_1, ..., X_n)$ corresponding to their DAG $\mathcal{G}$ by structural equations of the following form \cite{bollen2014structural}:
\[
X_i=f_i(\textbf{PA}(\mathcal{G},X_i),\epsilon_i)
\]
\noindent
where $\{\epsilon_i\}_{j=1,...,n}$ are mutually independent random noises and $\{f_i\}_{j=1,...,n}$ are real functions.

If a random vector ${X_1, ..., X_n}$ is generated according a SEM, we can factorize the density of the joint distribution as \cite{lauritzen1996graphical}:
\[
 p(x_1,...,x_n)=\prod_{1}^{n}p(x_i|\mathbf{PA}(\mathcal{G},x_i))
 \]
\noindent
It is clear that such a distribution is Markov to the DAG $\mathcal{G}$.

We now define an important concept called Pearl's \emph{do-intervention} \cite{pearl2009causality}. When we operate a do-intervention upon a variable $X_i$, we change the generating mechanism of $X_i$ and rewrite the SEM of $\mathbf{X}$ by updating the corresponding equation of $X_i$. This results in a new post-intervention distribution for 
$\mathbf{X}$. In particular, if the do-intervention fixes $X_i$ to a fixed point in the support of $X_i$, the joint density of $\mathbf{X}$ according to truncated factorization,
\[
 p(x_1,...,x_n|do(X_i=\hat{x_i}))=
  \begin{cases} 
   \prod \limits_{j\neq i} p(x_j|\mathbf{PA}(\mathcal{G},x_j)) & \text{if } X_i= x_i \\
   0      & \text{otherwise } 
  \end{cases}
\]

\textbf{Causal effects}
Let $(X,\mathbf{Y},\mathbf{U})$ be a random vector, where $\mathbf{Y}=(Y_1,...,Y_k)$. We define the total effects of $X$ on $\mathbf{Y}$ as
$$ (\mathbf{\tau}_{\mathbf{y}x})_i = \frac{d}{d(X)} \mathbb{E}(Y_i|do(X))$$ where $i \in (1,...,k)$

\textbf{Valid adjustment set} We call $\mathbf{Z}$ is a valid adjustment set for $(X,\mathbf{Y})$ if the following holds:
$$ p(\mathbf{Y})|do(X=x)= \int p(\mathbf{Y}|x,\mathbf{Z}=\mathbf{z})p(\mathbf{Z}=\mathbf{z}) d\mathbf{z}$$

\textbf{Total effects estimation via covariate adjustment} Let $\mathbf{\beta}_{\mathbf{st.w}}$ represent the least squares regression coefficient matrix whose $(i, j)$-th element is the regression coefficient of $T_j$ in the regression of $S_i$ on $\mathbf{T}$ and $\mathbf{W}$. Then the corresponding estimator $\hat{\mathbf{\beta}}_{\mathbf{y}x.\mathbf{z}}$ is an unbiased estimator for $\mathbf{\tau}_{\mathbf{y}x}$ while $\mathbf{Z}$ is a valid adjustment set for $(X,\mathbf{Y})$.

\textbf{Optimal adjustment set} Given $(\mathbf{X,Y})$, we denote the valid adjustment set whose total effect estimator attaining the smallest asymptotic variance among all valid adjustment sets as the optimal adjustment set, 

$$ \mathbf{O}_{\mathbf{x,y}}=\argmin_\mathbf{Z} \text{a.Var}(\hat{\mathbf{\beta}}_{\mathbf{yx}.\mathbf{z}})$$

\textbf{Bias and efficiency of a causal effect estimator} The bias of a causal effect estimator of $X$ on $\mathbf{Y}$, $\hat{\beta}$, is defined as $| \mathbb{E}(\hat{\beta})- \mathbf{\tau}_{\mathbf{y}x})|$. If $| \mathbb{E}(\hat{\beta})- \mathbf{\tau}_{\mathbf{y}x})|= \mathbf{0}$, we call $\hat{\beta}$ an unbiased estimator of $\mathbf{\tau}_{\mathbf{y}x})$. Then we could compare different \textit{unbiased} estimators by their variances. The smaller their variances are, the more efficient the estimators are referred as. By convention, as the estimators are generated via covariate adjustments, the concept of efficiency is also preserved when comparing different covariate sets for how precise they could generate causal effects estimators.

The possibility of multiple valid adjustment sets existing renders the subsequent selection of covariates an important and sometimes even imperative task, depending on various purposes such as dimension reduction or optimizing efficiency. Some approaches, such as those found in \cite{de2011covariate,glymour2008methodological,textor2011dagitty}, aim to select valid adjustment sets with minimal cardinality. Meanwhile, other approaches, such as \cite{li2005model,wang2012bayesian,shortreed2017outcome,talbot2015bayesian}, seek to minimize the variance of the causal effect estimator by including variables known as predictors of the outcome, which are statistically associated with the outcome, into the adjustment set.

The approaches aimed at obtaining valid adjustment sets with improved efficiency generally draw upon two key principles: conditional independence testing or outcome-based model selection. To be more specific, some methods, such as \cite{li2005model} or causal structure learning techniques (e.g. PC algorithm or Fast Causal Inference \cite{spirtes2000causation}) eliminate covariates from a given adjustment set if they are found to be conditionally independent of the outcome variable, given the exposure variable and other covariates, via parametric or nonparametric tests that align with the assumptions of the model. On the other hand, outcome-based model selection approaches such as those described in \cite{wang2012bayesian,shortreed2017outcome,talbot2015bayesian} strive to identify covariate sets that optimize metrics that assess the goodness of fit, such as AIC, BIC, p-value, or residual values.

In the realm of covariate selection, causal graphical information can prove to be a valuable asset. Through the graphical characterization of efficient valid adjustment sets, first explored in linear causal models by \cite{henckel2019graphical} and later expanded upon for non-parametric causal models by \cite{rotnitzky2019efficient}, one could determine the optimal adjustment set that has the lowest asymptotic variance among all valid adjustment sets, based on the true causal Directed Acyclic Graph (DAG). Unfortunately, the true causal graph is often unavailable, rendering this graphical characterization inapplicable. As for causal graphs that have been estimated from observational data, it is necessary to assume causal sufficiency before using the results from \cite{henckel2019graphical,rotnitzky2019efficient}, which is not only unrealistic, but also raises concerns regarding the accuracy and credibility of the causal structure estimation. 

\section{Performance of covariate selection methods for causal inference with hidden variables}

Firstly, our research focuses on a specific group of covariate selection methods that rely on fitting a model of the outcome variables to gauge its predictive capability and thereby prune an efficient adjustment set from a larger one. (The other group has its own limitations which are out the scope of this paper.) By forgoing the requirement of causal sufficiency, we examine and evaluate their performance in causal inference with hidden variables. This allows us to provide a definitive answer to the intriguing question of whether these methods are still capable of producing valid adjustment sets in this section. 

\subsection{Examination of validity as adjustment sets} 
It is common practice in prediction-based approaches to use least square methods, such as OLS, LASSO, or similar, to prune variables from a given valid adjustment set that are considered to be "predictors" of the outcome variable in order to create a more efficient valid covariate set. While this approach may result in causal effect estimators with lower variance, it is not necessarily valid and can introduce biases when strong premises such as causal sufficiency are violated. The problem arises from the difference between what should be the "predictors" of the outcome and the variables that are selected through these approaches. Simply having covariates with zero coefficients in a regression model against a dependent variable does not indicate conditional independence between the variables given the remaining covariates. Unfortunately, many studies either intentionally or unintentionally overlook this important distinction.

In this subsection, we shall formally define what is considered as the "predictors" of the outcome in many prediction-based approaches and determine the actual set of variables that are obtained through these methods within the context of causal graphs.

We consider the following problem setting and define key terms that will be used later in the study. Let $\mathcal{G}$ be a causal DAG with nodes set $\mathbf{V}$ containing $X,Y$ and only non-descendants of $X,Y$. Suppose $\mathbf{Z} \subseteq \mathbf{V}$ is a valid adjustment set relative to $X,Y$ in $\mathcal{G}$ and distribution of $\mathbf{V}$ is faithful to $\mathcal{G}$. For simplicity, we restrict our attention for situations when $\mathbf{V}$ admits a linear additive model.(Extension to general additive models is possible.) Let $\mathbf{T}(X,Y)$ denote $\argmin_{Z_i:\beta_i\neq 0}\mathbb{E}(Y- \alpha X-\mathbf{Z}\mathbf{\beta})^2$.

\begin{Definition}
\normalfont
Predictors of $Y$ are all the $Z \in \mathbf{Z}$ such that $Z \notindependent Y |X\cup \mathbf{Z} \setminus Z$
\end{Definition}

To illustrate definition of predictors, we shall refer to Figure 1 for different configurations of $\mathbf{Z}$ and corresponding predictors of $Y$ as examples. 
\begin{itemize}
    \item If $\mathbf{Z}=\{Z_1,Z_2,Z_3\}$, the predictors of $Y$ is $Z_2$.
    \item If $\mathbf{Z}=\{Z_1,Z_3\}$, the predictors of $Y$ are $\{Z_1,Z_3\}$.
    \item  If  $\mathbf{Z}=\{Z_2,Z_3\}$, the predictor of $Y$ is $Z_2$. 
\end{itemize}

\begin{Definition}
\normalfont 
Two nodes in $Z_1,Z_2 \in X\cup \mathbf{Z}$ are d-adjacent if there exists a path between $Z_1,Z_2$ on $\mathcal{G}$ that is not blocked by $\mathbf{Z}\setminus \{Z_1,Z_2\}$
\end{Definition}

It is clear that any two adjacent nodes are d-adjacent. It is also important to note that two non-adjacent nodes of a graph may become d-adjacent when considering only a subset of the nodes, which is a common scenario in models with hidden variables.  For instances, in Figure 1, both $Z_1,Z_3$ are d-adjacent to $Y$ in the set $\{X,Z_1,Z_3,Y\}$.

\begin{figure}
    \centering
        \begin{tikzpicture}[main/.style = {draw, circle}] 
           \node[main] (1) {$X$};
           \node[main] (2) [above right of=1] {$Z_1$};
           \node[main] (3) [right of=2] {$Z_2$};
           \node[main] (4) [right of=3] {$Z_3$};
           \node[main] (5) [below right of=4] {$Y$}; 
           \draw[->] (2) -- (1);
           \draw[->] (1) -- (5);
           \draw[->] (2) -- (3);
           \draw[->] (3) -- (4);
           \draw[->] (3) -- (5);

        \end{tikzpicture}     
    \caption{Example illustrating Proposition 1(c)}
    \label{fig:my_label}
\end{figure}
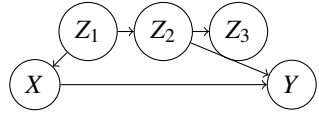

Proposition 1 underscores the crucial difference between the outcome predictors and $\mathbf{T}(X,Y)$ in terms of their graphical representation and their validity as adjustment sets. It is evident that when $\mathbf{Z}$, a valid adjustment set relative to $X,Y$, is given, the predictors of $Y$ also form a valid adjustment set. However, this is not always the case for $\mathbf{T}(X,Y)$, as demonstrated in Proposition 1. 

\begin{Proposition}
\normalfont
Let $\mathcal{G}$ be a causal DAG with nodes set $\mathbf{V}$ containing $X,Y$ and only non-descendants of $X,Y$. Suppose $\mathbf{Z} \subseteq \mathbf{V}$ is a valid adjustment set relative to $X,Y$ in $\mathcal{G}$. Then 
\begin{enumerate}[label=\alph*]
    \item Predictors of $Y$ are exactly all variables in $X\cup \mathbf{Z} $ that are d-adjacent to $Y$.

    \item If $Y \independent \mathbf{Z} \setminus \mathbf{T}(X,Y)|X \cup \mathbf{T}(X,Y)$, then $\mathbf{T}(X,Y)$ is a valid adjustment set relative to $X,Y$ in $\mathcal{G}$ and constitutes the optimal adjustment in $\mathbf{Z}$. 
    \item  $\nexists W \in \mathbf{V} $ such that there exist two variables in predictors of $Y$ in $\mathbf{Z}$ that are not connected with $Y$ in $\mathbf{Z} \cup W $, then $\mathbf{T}(X,Y)$ is exactly the predictors of $Y$ and a valid adjustment set.
\end{enumerate}

\end{Proposition}

We give an example illustrating Proposition 1(c) according to Figure 1. Assuming all coefficients in the underlying linear structural causal model are all 1's with $Z_2$ missing, we can deduce the following results:

\begin{itemize}
    \item If $Var(Z_3) < Var(Z_1)$, then $\mathbf{T}(X,Y)=Z_3$
    \item If $Var(Z_3) > Var(Z_1)$, then $\mathbf{T}(X,Y)=Z_1$
\end{itemize}
However, it is clear that only $Z_1$ is a valid adjustment set relative to $X,Y$ but not $Z_3$.

Thus far, our analysis has shown that when the underlying graphical structure is unknown, the results of prediction-based methods may deviate from the true outcome predictors, and more significantly, may not remain valid as a covariate adjustment set in the presence of hidden variables. It is dangerous to draw causal inferences from these results or to employ them for causal inquiries such as interventions.

\subsection{Predictive power revisit}

While extra precaution is advised when using prediction-based covariate selection methods to extract causal information, it is not surprising to find that these methods maintain their eponymous characteristics in terms of predictive performance even with hidden variables present.

\begin{Proposition}
 \normalfont
 Let $\mathcal{G}$ be a causal DAG with nodes set $\mathbf{V}$ containing $X,Y$ and only non-descendants of $X,Y$. Suppose $\mathbf{Z} \subseteq \mathbf{V}$ is a valid adjustment set relative to $X,Y$ in $\mathcal{G}$, following a linear causal model. Let $\mathbf{T}(X,Y)$ denote $\argmin_{Z_i:\beta_i\neq 0}\mathbb{E}(Y- \alpha X-\mathbf{Z}\mathbf{\beta})^2$. Then, $\mathbb{E}(\hat{\mathbf{\beta}}_{\mathbf{y}x.\mathbf{T}}) = \tau_{yx}$ at and $\textbf{var}(\hat{\mathbf{\beta}}_{\mathbf{y}x.\mathbf{T}}) \leq \textbf{var}(\hat{\mathbf{\beta}}_{\mathbf{y}x.\mathbf{Z}})$ at large sample.
\end{Proposition}

 The key aspect to Proposition 2 is when the underlying data generating mechanism is assumed to be a linear causal structural model, with the potential presence of hidden variables, $\mathbf{T}(X,Y)$ can provide a numerically accurate estimate for causal effect estimation. However, this assertion cannot be extended to non-parametric models or in applications such as domain adaptation, where a purely observational model may not be adequate.

\section{Causal Mixed Integer Optimization algorithm for optimal valid covariate selection}

As previously noted, the use of prediction-based covariate selection methods can lead to more efficient causal estimates under the assumption of causal sufficiency. In this section, we will delve deeper into this important subject and explore ways to further improve the efficiency of causal effect estimators by constructing novel approach to select valid adjustment set.

We present a novel approach for pruning the valid covariate adjustment set, which has the minimum asymptotic variance of its corresponding causal effect estimator from any given set of observed covariates that are valid under certain mild conditions outlined below. Our method has several key features that set it apart from existing approaches. Firstly, it is a non-graphical approach that can be applied to observational data. Secondly, it focuses on exploring the implicit causal connotation of variables in an optimization model, rather than the predictive accuracy, as is often the case with state-of-the-art methods. Thirdly, we have proven its asymptotic consistency in achieving the optimal covariate adjustment set and have shown its superiority in a variety of data generating schemes with finite samples, including high-dimensional ones, in terms of both optimal covariate set identification and subsequent causal effect estimation.

Before presenting our own methodology, it is imperative that we first introduce a renowned work on best subset selection, which forms the basis our investigation.

\subsection{ Mixed Integer Optimization algorithm}

It is widely recognized that the classical approach of best subset selection in a linear regression model with a constraint on the cardinality of non-zero regression coefficients has limitations in scaling to a large number of covariates. However, this changed with the introduction of the mixed integer optimization (MIO) algorithmic framework by \cite{bertsimas2016best}. The remarkable advancements in the capabilities of MIO solvers, such as Gurobi and Cplex, now allow the methods proposed in \cite{bertsimas2016best} to provide solutions to best subset problems with thousands of covariates in mere minutes, making it both practical and appealing. For further information on the properties of the algorithmic framework and its theoretical and simulation results in the context of linear regression, we direct readers to consult \cite{bertsimas2016best}. In what follows, our focus shifts to incorporating the ideas behind MIO solvers for best subset selection into the realm of efficient covariate selection for causal estimation, a distinct but related challenge.

\subsection{Mixed Integer Optimization Formulations for optimal covariate set selection problem}

Throughout this section,  the problem settings as well as assumptions are given as below. Given a dataset of $n$ observations derived from a random vector $X\cup Y \cup \mathbf{Z}$, which is generated in accordance with a linear structural equation model\cite{pearl2009causality}, our objective is to determine the optimal covariate adjustment set associated with the variables $X$ and $Y$ within $\mathbf{Z}$, under the assumption that
\begin{enumerate}
    \item $\forall Z \in \mathbf{Z} $ is a nondescendant of $X,Y$.
    \item $\mathbf{Z} $ is a valid adjustment set relative to $X,Y$.
    \item There are no hidden variables in $\mathbf{Z} $.
    
\end{enumerate}

The first two assumptions, referred to as pretreatment covariates and conditional exchangeability, are widely acknowledged in the causal inference literature \cite{witte2019covariate}. The third assumption, causal sufficiency, is implicitly assumed in most of the covariate selection methods currently under examination of our paper. 

Thanks to graphical criteria put forth by \cite{henckel2019graphical}, we are able to determine the composition of the optimal adjustment set, providing insight into what we should be searching for even in the absence of a causal graph. 

\begin{Proposition}
\normalfont
Given the problem setting and assumptions 1-3, then the optimal covariate adjustment set of $X,Y$, $\textbf{O}(X,Y)$, is $\mathbf{PA}(\mathcal{G},Y)\setminus X$.

\end{Proposition}

With the aim of attaining the optimal adjustment set outlined in Proposition 3, we present an adaptation of the Mixed Integer Optimization problem incorporating additional causal constraints as stated in equation (\ref{eqn1}). To this end, we also propose a corresponding optimization algorithm capable of yielding a set that can be proven to be $\mathbf{PA}(\mathcal{G},Y)$ in probability. 

\begin{equation}
\label{eqn1}
\begin{aligned}
\max_{k}\min_{\mathbf{\beta},\mathbf{u}} \quad & \frac{1}{2}||Y- \alpha X-\mathbf{Z}\mathbf{\beta}||^2_2\\
\textrm{s.t.} \quad & -Cu_i\leq \beta_i\leq Cu_i;u_i\in \{0,1\};\sum_i^p u_i=k; i=1,...,p\\
  & Y \notindependent Z_i|A, \forall A \in \mathbf{Z}^+, \textrm{where} \beta_i \neq 0, \mathbf{Z}^+=\{Z_i:\beta_i\neq 0\}\\
\end{aligned}
\end{equation}

\begin{Proposition}
\normalfont

Solution to (\ref{eqn1}) exists uniquely and is $\mathbf{PA}(\mathcal{G},Y)$ in probability.

\end{Proposition}

Inspired by "Algorithm 1" in \cite{bertsimas2016best}, we present an algorithm that provides a solution to (\ref{eqn1}) by incorporating the existing MIO algorithm by \cite{bertsimas2016best} with the general properties of conditional independence imposed by the parents of $Y$. We refer to this algorithm as CMIO. The convergence properties of this algorithm are discussed in Section 4.3. 

It is worth noticing that, empirically, the use of d-separation relationships in CMIO may be substituted with a measure of conditional independence, such as the p-value of t-tests for linear models with Gaussian noise, or any other nonparametric implementation of conditional independence testing.

\begin{algorithm}[H]
\caption{}
\label{alg:euclid}
\begin{algorithmic}[1]
\State \textbf{INPUT}:Given a dataset of $\mathbf{Z}$, $k=1$,$\mathbf{O}=\emptyset$

    \While{$k\leq p$}
    
    \State{ Employ MIO algorithm with $k$-sparsity to solve $\min ||Y-\beta (X,\mathbf{Z}^T)^T||_2^2$. Denote the solution as $\mathbf{Z^k}$  where $\mathbf{Z}^{0}=\emptyset$}
    \If{$\mathbf{Z}^{k-1} \subseteq \mathbf{Z}^k$ }{$Z^{MIO}_k=\mathbf{Z}^k \setminus \mathbf{Z}^{k-1}$}
    \If{$Z^{MIO}_k \notindependent Y |A, \forall A\subseteq \mathbf{Z}^{k-1}$} { $\mathbf{O}=\mathbf{Z^k}$, $k=k+1$}

    \Else{ $\mathbf{O}=\mathbf{Z^{k-1}}$, $k=p+1$ }
    \EndIf
    \Else{ $\mathbf{O}=\mathbf{Z^k}$ $ k=k+1$}
    \EndIf
    \EndWhile
    
    \Return{$\mathbf{O}$}

\end{algorithmic}
\end{algorithm}

\subsection{Statistical properties of CMIO algorithm in the causal context}
In their study, \cite{bertsimas2016best} has made a significant contribution by outlining the theoretical properties, including the convergence properties, of the MIO algorithm. We recommend adopting their findings as a basis when discussing the statistical properties of CMIO as they are blosed related. For instance, the results obtained from the MIO algorithm are considered to be exact solutions to the optimization problems, subject to the constraints specified in the algorithm. In the following theorem, we present the theoretical guarantees that underpin the validity of our approach, the CMIO.

\begin{Theorem}
\normalfont 
Algorithm 1 estimates the optimal covariate set related to $X$ on $Y$ with probability 1 as $n \rightarrow \infty$. 
\end{Theorem}

In the appendix, we present a proof of Theorem 1 which demonstrates the soundness of Algorithm 1 in identifying the optimal covariate set. This is accomplished through the use of an adapted procedure from a MIO solver, as recommended in \cite{bertsimas2016best}. In this approach, the properties of MIO solutions are harnessed to address a similar optimization problem that includes causal constraints. 

\section{CMIO with hidden variables}
In the preceding section, we have primarily focused on the application of the MIO solver in facilitating efficient estimation of causal effects, under the assumption of causal sufficiency. A logical next step would be to examine the scenario where this assumption is no longer valid. 

With the presence of hidden variables, neither the optimal covariate set nor the predictors of outcome is identifiable solely from observed variables. And the  optimization solution $\mathbf{T}(X,Y)$ has been proved to be possibly an invalid adjustment set. As a result, it becomes imperative to find new and innovative ways to address this issue. Fortunately, we have discovered that adding certain variables in $\mathbf{Z}$ to $\mathbf{T}(X,Y)$ helps resolve the problem. 

\begin{Proposition}
\normalfont
Let $\mathcal{G}$ be a causal DAG with nodes set $\mathbf{V}$ containing $X,Y$ and only non-descendants of $X,Y$. Suppose $\mathbf{Z} \subseteq \mathbf{V}$ is a valid adjustment set relative to $X,Y$ in $\mathcal{G}$. Then we can select $\mathbf{Z}'$, a subset of $\mathbf{Z}$ that is valid covariate adjustment set relative to $X,Y$ in $\mathcal{G}$ and more efficient than $\mathbf{Z}$ for causal effect estimation. Explicitly, $\mathbf{Z}'$ is comprised of the following: $\mathbf{Z}'= \mathbf{T}(X,Y) \cup \mathbf{S}$, where $\mathbf{S}=\{S\in \mathbf{Z},: S \notindependent Y |\mathbf{T}(X,Y) \cup X\}$. 
\end{Proposition}

Proposition 5 has enabled us to formulate a target set in a specific way, resulting in a valid adjustment set that coincides with predictors of the outcome. This is the optimal result that can be achieved using non-graphical methods. However, it cannot be guaranteed that this set is the optimal adjustment set in $\mathbf{Z}$. It is also untestable whether removing a variable that is d-adjacent to $Y$ and connected with $X$ from a valid adjustment set would result in an invalid set, unless the assumption of causal sufficiency is made. This issue has been previously documented in literature, as evidenced by \cite{henckel2019graphical}.

Algorithm 2 represents an implementation of the procedure outlined in Proposition 5, which serves as a complement to Algorithm 1 for dealing with covariate selection in the presence of hidden variables. The first part of the iteration in Algorithm 2 is dedicated to identifying variables that are conditionally dependent on $Y$ given $X \cup \mathbf{O}$ in the remaining set $\mathbf{Z}\setminus \mathbf{O}$. The procedure successfully finds a set that complements $\mathbf{O}$ and forms a valid adjustment set. In order to minimize the size of the complement set, a second iteration is conducted, which resembles a backward selection process to remove any redundant variables. Under the assumption of a linear model with Gaussian errors, conditional independence testing can be performed using Fisher's Z-transform at a chosen significance level. Alternatively, other measures of conditional independence can be used, as demonstrated in \cite{li2005model} through a non-parametric testing approach, to accommodate various data distributions.

\begin{algorithm}[H]
\caption{}
\label{alg:euclid}
\begin{algorithmic}[1]
\State \textbf{INPUT}: $\mathbf{O}$, which is the output of Algorithm 1,
a data set of $\mathbf{Z}$
\State{$\mathbf{T}= \mathbf{Z} \setminus \mathbf{O}$ }
\For{ $\forall T \in \mathbf{T} $ }
\If{$T \independent Y | X \cup \mathbf{O}$}{$\mathbf{T}=\mathbf{T} \setminus T$}
\EndIf
\EndFor
\For{ $\forall T \in \mathbf{T} $ }
\If{$T \independent Y | X\cup \mathbf{O}\cup \mathbf{T}\setminus T$}{$\mathbf{T}=\mathbf{T} \setminus T$}
\EndIf
\EndFor

    \Return{$\mathbf{O}\cup \mathbf{T}$}

\end{algorithmic}
\end{algorithm}

Once again, we refer back to Figure 1 and illustrate the discrepancy of existing methods and our revised CMIO on data generated from linear structural model with $Z_2$ missing. Assuming all coefficients in the underlying linear structural causal model are all 1’s, and $Var(Z_3) < Var(Z_1)$, then $\mathbf{T}(X,Y)=Z_3$, which is clearly invalid as covariate set, while Algorithm 2 would produce $\{Z_1,Z_3\}$, a valid adjustment set over $X,Y$. 

\section{Simulation study}
In this section, we evaluate the performance of the proposed CMIO algorithm in comparison to established state-of-the-art methods using simulated data. The simulations, based on illustrations presented by \cite{ertefaie2018variable} and \cite{shortreed2017outcome}, consist of multivariate Gaussian distributed covariates $\mathbf{Z}$, binary $X$, and continuous $Y$, generated through logit and linear regression, respectively. We consider three different data sets with varying correlation levels of $\mathbf{Z}$ and sample sizes. The first two scenarios involve low-dimensional data with 100 covariates, 20 of which are parents of $Y$, and sample sizes of either $n=200$ or $n=1000$. The third scenario involves high-dimensional data with 100 covariates and $|\mathbf{Pa}(Y)|=4$, where $|\mathbf{Z}|=100>n=50$. Each scenario was run 100 times.

The data generating processes of three cases are specified as follows:

\begin{enumerate}
    \item[Case 1.] $\mathbf{Z}=(Z_1,...,Z_{100})^T \sim \mathcal{N}(\mathbf{0},I_{100})$ $X \sim  \text{Bern}(p)$, where $p=\text{logit}(Z_1+Z_2+...+Z_{10}+Z_{21}+Z_{22}+...+Z_{30})$ $Y=0.5*X+0.6*(\sum_1^{20} Z_i)+\epsilon$, where$\epsilon \sim \mathcal{N}(0,1)$
    
    \item[Case 2.] $\mathbf{Z}=(Z_1,...,Z_{100})^T \sim \mathcal{N}(\mathbf{0},\Sigma)$, where $\Sigma_{ij}=1$ $\text{ if }i=j$ $\Sigma_{ij}=0.5$ $\text{ if }i \neq j$  $X \sim  \text{Bern}(p)$, where $p=\text{logit}(Z_1+Z_2+...+Z_{10}+Z_{21}+Z_{22}+...+Z_{30})$ $Y=0.5*X+0.6*(\sum_1^{20} Z_i)+\epsilon$, where$\epsilon \sim \mathcal{N}(0,1)$
    
    \item[Case 3.] $\mathbf{Z}=(Z_1,...,Z_{100})^T \sim \mathcal{N}(\mathbf{0},\Sigma)$, where $\Sigma_{ij}=1$ $\text{ if }i=j$ $\Sigma_{ij}=0.5$ $\text{ if }i \neq j$  $X \sim  \text{Bern}(p)$, where $p=\text{logit}(0.5*Z_1-0.5*Z_2+0.3*Z_5-0.3*Z_6+0.35*Z_7+0.4*Z_8)$ $Y=X+2*(\sum_1^{4} Z_i)+\epsilon$, where$\epsilon \sim \mathcal{N}(0,1)$
\end{enumerate}

In this section, we conduct a comparative evaluation of our proposed CMIO algorithm against two benchmark methods: the Bayesian Causal Effect Estimation (BCEE) method proposed by Talbot et al. (2015) and the Outcome Adaptive Lasso (OLA) method put forward by Shortreed et al. (2017). We also compare CMIO's performance to the oracle target set $\mathbf{PA}(\mathcal{G},Y)$.

Other well-known methods such as Bayesian Model Averaging proposed by Wang et al. (2012) and a Non-Parametric method by De and Bühlmann (2011) have been shown to perform poorly compared to BCEE and OLA in previous studies (Shortreed et al., 2017), and thus we do not consider them in our evaluation.

We evaluate CMIO's performance in two dimensions: the precision of the estimated causal effect and the ability to discover the optimal adjustment set. To assess the precision of the estimated causal effect, we present the results in boxplots using two metrics: the set difference between the estimated and actual covariate set, and the percentage of the estimation being valid and containing all members of the target set.

Figures 2 and 3 present box plots of estimated causal effects obtained from various methods as part of our simulation studies for Case 1 and Case 2 scenarios, respectively. Across both scenarios, our proposed method CMIO performs exceptionally well, closely approximating the results of the target set. Moreover, even at a small sample size of 200 observations, CMIO achieves an overall mean value that is closely aligned with the true causal effects, outperforming both BCEE and OLA on both objectives. As we have deliberately selected the most advanced existing methods for comparison, it is not unexpected that all methods perform satisfactorily in terms of accuracy at a larger sample size.

Regarding high-dimensional data, we have summarised our findings in the box plot presented in Figure 4. OLA was excluded from this analysis as it does not support high-dimensional data. The plot clearly indicates that CMIO excels in estimation accuracy, performing significantly better than BCEE and following the target set quite closely without deviating substantially. 

It is essential to examine the validity of CMIO’s output as an adjustment set, which can be evaluated by including all confounding variables. Therefore, we have analysed the difference between the estimated and optimal adjustment sets produced by CMIO, as well as the proportion of instances where all confounders are included in CMIO’s estimates. The CMIO algorithm imposes a cardinality constraint, so only a few spurious variables are selected, if any. With the exception of scenarios with highly correlated data for confounders with a smaller sample size of 200, CMIO consistently identifies a valid adjustment set that includes all confounders. This reassures us of the unbiasedness of the causal effects estimations produced by CMIO apart from the boxplots in previous graphs.

Overall, our results show that CMIO performs well in all simulated cases, delivering a perfect selection of the optimal adjustment set at $n=1000$ while maintaining high validity (95$\%$) even for small sample sizes. CMIO also produces near-perfect results when the data is high-dimensional and comprised of moderately correlated covariate variables, demonstrating the superior efficacy of our proposed CMIO algorithm in efficient covariate selection.

\begin{figure}
    \centering
    \begin{subfigure}[b]{0.45\textwidth}
      \includegraphics[width=\textwidth]{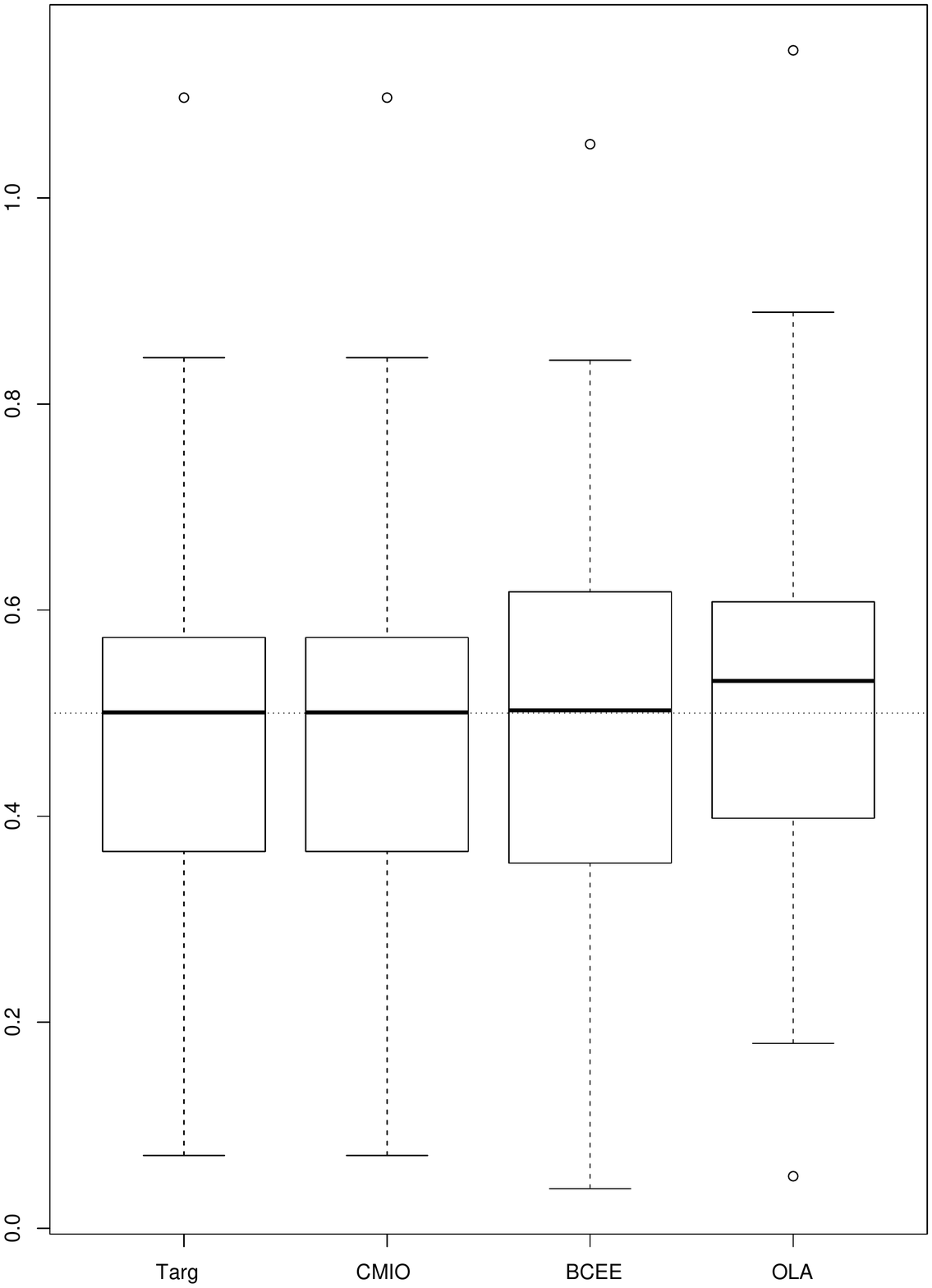}
    \caption{Case 1, n=200}

    \label{fig:my_label}

    \end{subfigure}
    \begin{subfigure}[b]{0.45\textwidth}
    \includegraphics[width=\textwidth]{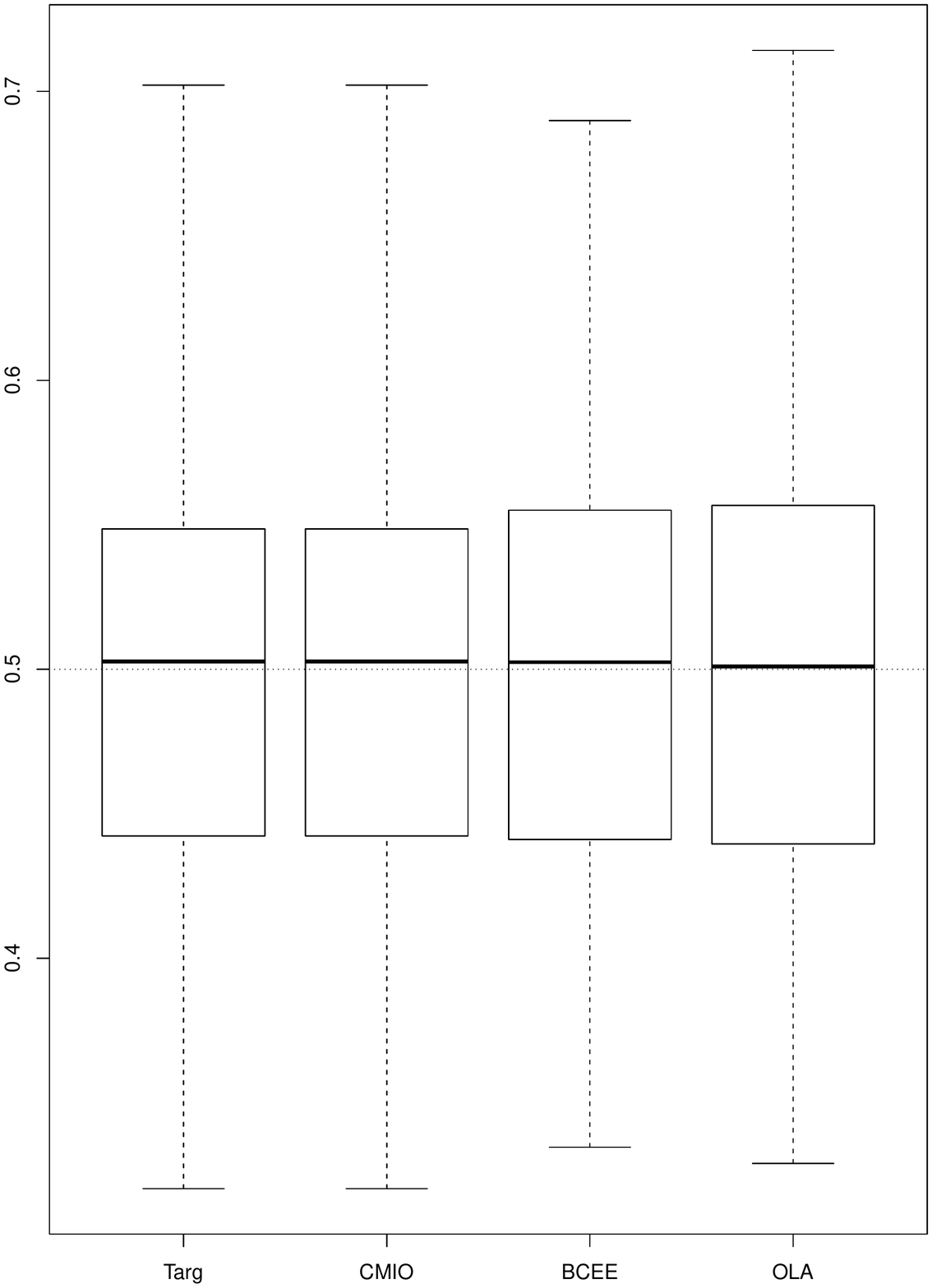}
    \caption{Case 1, n=1000}
    \end{subfigure}
\caption{Boxplot of estimated causal effects for case 1. Horizontal dashed line indicate the true treatment effect.}
\end{figure}

\begin{figure}
    \centering
    \begin{subfigure}[b]{0.45\textwidth}
      \includegraphics[width=\textwidth]{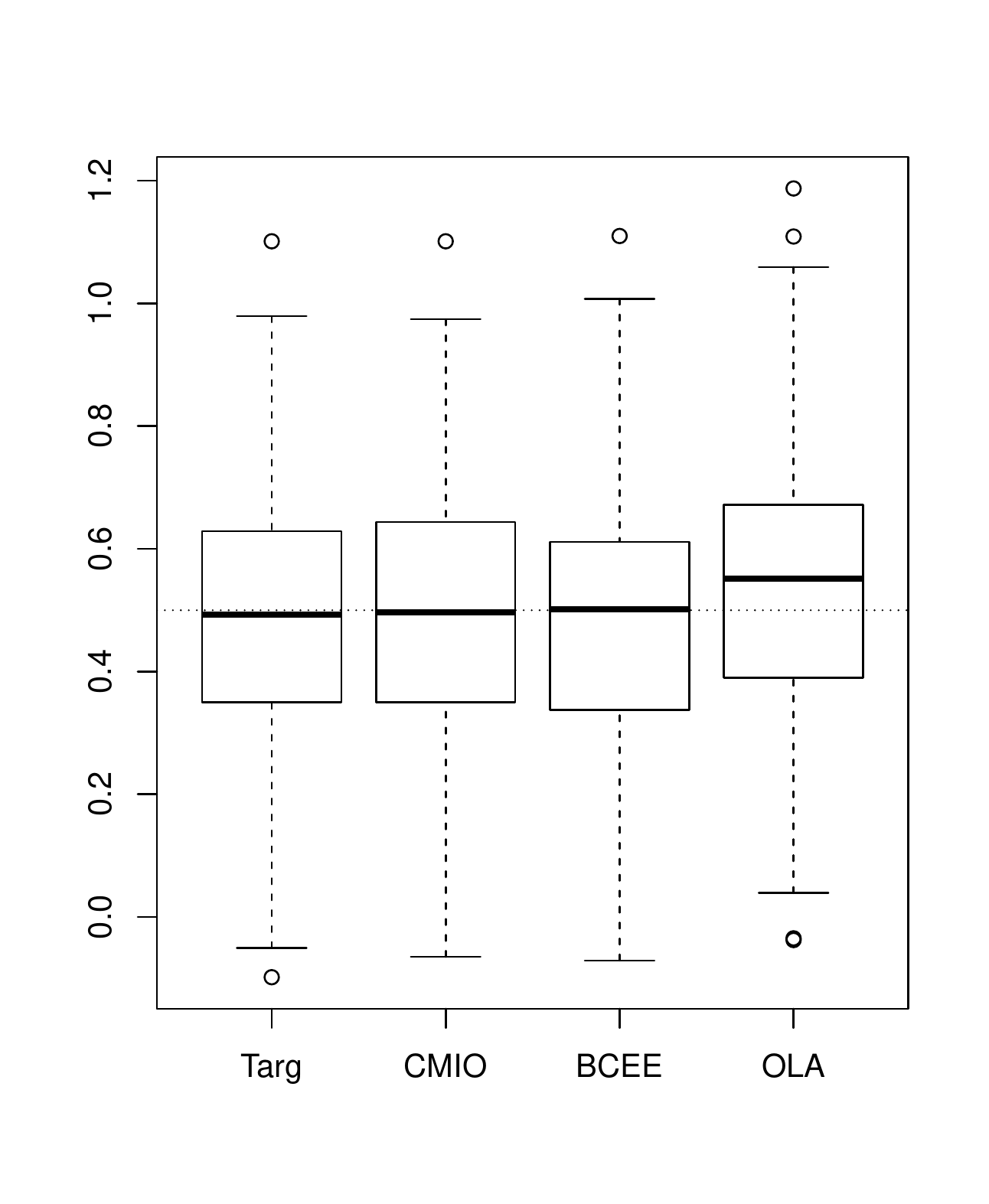}
    \caption{Case 2, n=200}

    \label{fig:my_label}

    \end{subfigure}
    \begin{subfigure}[b]{0.45\textwidth}
    \includegraphics[width=\textwidth]{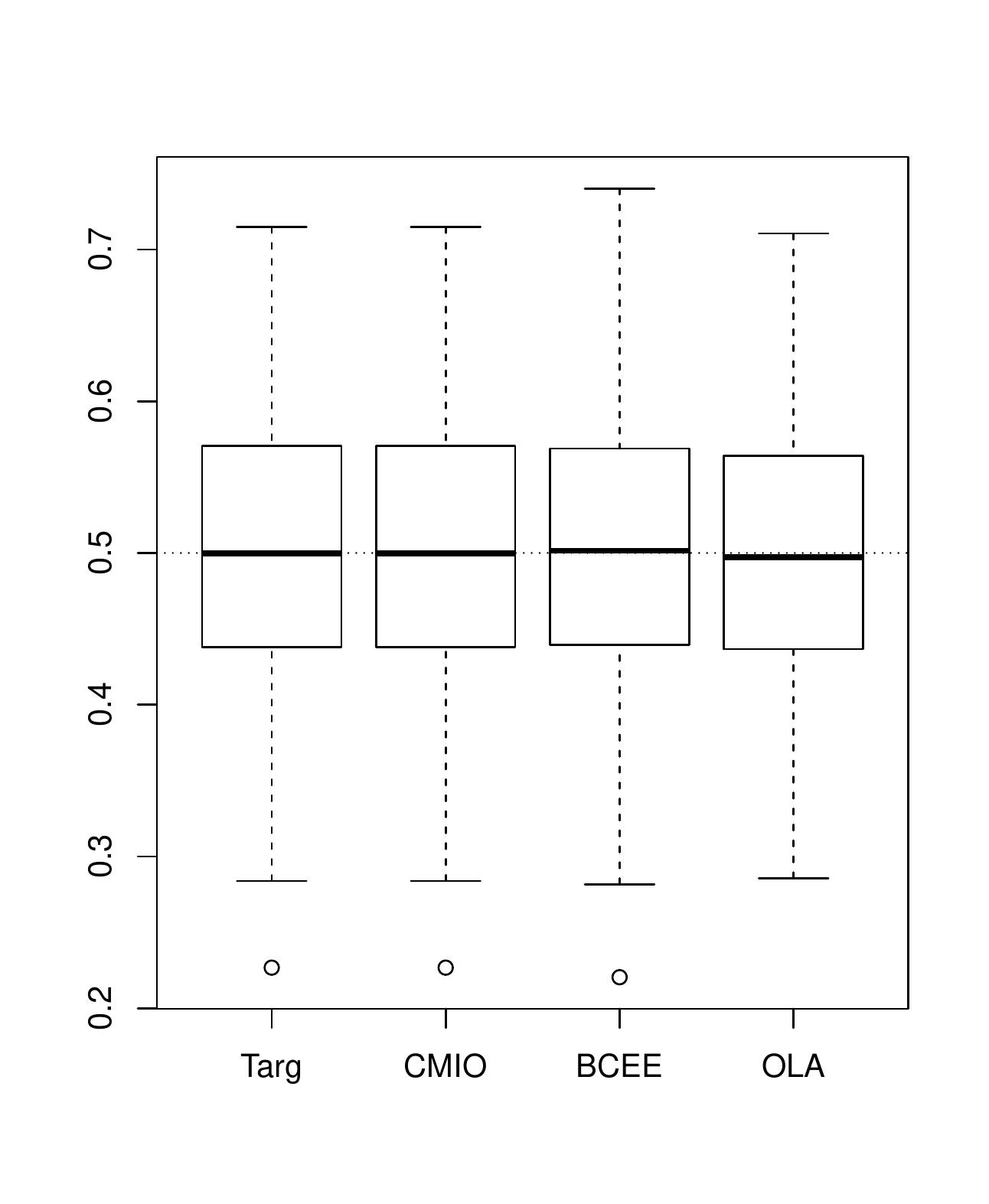}
    \caption{Case 2, n=1000}
    \end{subfigure}
\caption{Boxplot of estimated causal effects for case 2. Horizontal dashed line indicates the true treatment effect.}
\end{figure}

\begin{figure}
    \centering
    \includegraphics[width=0.5\textwidth]{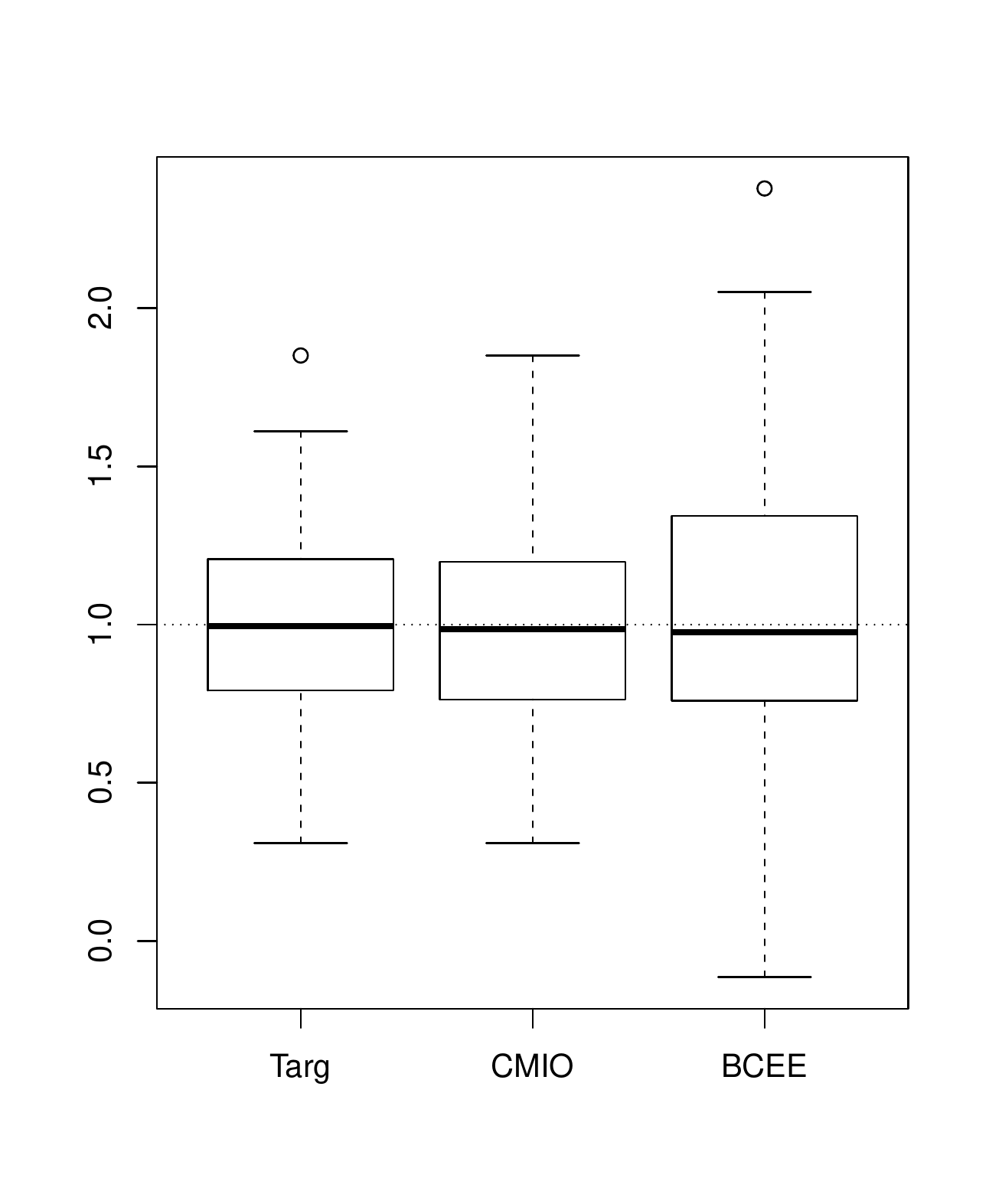}
    \caption{Boxplot of estimated causal effects for case 3. Horizontal dashed line indicates the true treatment effect.}
    \label{fig:my_label}
\end{figure}
\begin{table}[h]
    \centering
    \begin{tabular}{p{0.30\textwidth} p{0.30\textwidth}p{0.30\textwidth}}
      \hline
       & Size of Set difference & $\% \mathbf{Pa}(Y) \subseteq \hat{\mathbf{O}} $ \\
      \hline\hline
      Case 1, n=200 & 0.12 $\pm$ 0.42 &100\\
      Case 1, n=1000 & 0  & 100  \\
      Case 2, n=200 & 0.96 $\pm$ 0.83  & 95  \\
      Case 2, n=1000 & 0  & 100  \\
      Case 3, n=50 & 0.06 $\pm$ 0.24  & 100 \\
      \hline
    \end{tabular}

    \caption{Summary of set difference between the optimal adjustment set and the estimated ones by CMIO and proportion of time the estimated covariate set containinng the target set.}
    \label{tab:my_label}
\end{table}

\section{Discussion}
In this paper, we provide insights into field of covariate selection methods in causal inference, focusing on the validity and efficiency of these methods. Specifically, the study examines the properties of existing prediction-based covariate selection methods and caveats about the potential dangers of producing invalid adjustment sets in the presence of hidden variables and lack of graphical information.

In response to this, the study proposed a novel graph-free method, CMIO, which enhances the efficiency of existing methods. The method is based on Mixed Integer Optimization and equipped with a set of causal constraints, ensuring that the output is theoretically sound. The study compares the performance of CMIO with other state-of-the-art methods and shows that CMIO outperforms them in terms of validity and efficiency.

Finally, the study revisits the scenario in the absence of causal sufficiency causal graphs and presents a revised version of CMIO capable of handling such circumstances. The statistical properties of this revised method are rigorously demonstrated and offer a promising solution for efficient and valid covariate selection in causal inference.

\newpage

\appendix
\section{}
\label{app:theorem}



In this appendix we prove propositions and theorems stated in the paper.
\subsection{Proof of Proposition 1}
\textit{Proof}
a. Let $Z \in X\cup \mathbf{Z}$ that is connected to $Y$. By Definition 2,we have $Z \notindependent_{\mathcal{G}} Y |X\cup \mathbf{Z} \setminus Z$. From faithfulness assumption, we can deduce that $Z \notindependent Y |X\cup \mathbf{Z} \setminus Z$, which shows that $Z$ is a predictor from Definition 1. Similarly, let $Z \in X\cup \mathbf{Z}$ be a predictor of $Y$, then $Z \notindependent Y |X\cup \mathbf{Z} \setminus Z$ and by Markovian property, $Z \notindependent_{\mathcal{G}} Y |X\cup \mathbf{Z} \setminus Z$, which means $Z$ is connected to $Y$ in $\mathbf{Z}$ by Definition 2. \qedsymbol 

b. According to Theorem 3.4.1 of \cite{pearl2009causality}, $\mathbf{T}(X,Y)$ is a valid adjustment set relative to $X,Y$ in $\mathcal{G}$.  Suppose that we have two valid adjustment sets $\mathbf{T}(X,Y)$ and $\mathbf{Z}_1$, where $\mathbf{Z}_1 \in \mathbf{Z} $, then clearly it can be deduced from $Y \independent \mathbf{Z} \setminus \mathbf{T}(X,Y)|X\cup \mathbf{T}(X,Y)$ that $Y \independent \mathbf{Z}_1 \setminus \mathbf{T}(X,Y)|X\cup \mathbf{T}(X,Y)$ since $\mathbf{Z}_1 \in \mathbf{Z}$. On the other hand,  $\forall S \in \mathbf{Z}_1 \setminus \mathbf{T}(X,Y)$ it must satisfy wither $S \independent X |\mathbf{Z}_1$ or $S \independent Y |\mathbf{Z}_1$ as $\mathbf{Z}_1$ being a valid adjustment set. Furthermore, $S \notindependent Y |\mathbf{Z}_1$ due to $S \in \mathbf{T}(X,Y)$, indicating $S$ is d-adjacent to $Y$. Hence, we must have $S \independent X |\mathbf{Z}_1$. By Theorem 3.4 of \cite{henckel2019graphical}, combining the two conditions above, we can obtain that the asymptotic variance of causal effect estimators satisfy :$\textbf{var}(\hat{\mathbf{\beta}}_{\mathbf{y}x.\mathbf{Z}_1}) \ge \textbf{var}(\hat{\mathbf{\beta}}_{\mathbf{y}x.\mathbf{T}})$ \qedsymbol

c. First, we notice that $\mathbf{T}(X,Y)$ is a subset of predictors of $Y$. Suppose that $\exists U\in $ predictors of $Y$ with $U \notin \mathbf{T}(X,Y)$, by uniqueness of the least squares regression, $U$ is not a parent of $Y$ in $\mathcal{G}$, otherwise its coefficient would be non-zero and belongs to $\mathbf{T}(X,Y)$. Hence, we can find a subset of parents of $Y$ in $\mathbf{V}$, say $\mathbf{W}$ such that $Y \independent U|X \cup \mathbf{W}$. Clearly, $\exists T \in \mathbf{T}(X,Y), T \notin \mathbf{W}$ that is d-separated with $Y$ by some of $\mathbf{W}$, contradicting our assumption that no two members of the predictors of $Y$ are not connected in $\mathbf{Z}\cup \mathbf{W}$. Therefore, every member of predictors of $Y$ belongs to $\mathbf{T}(X,Y)$. Hence, predictors of $Y$ are exactly those of  $\mathbf{T}(X,Y)$ and are valid. \qedsymbol

\subsection{Proof of Proposition 2}
\textit{Proof} 
Based on the uniqueness of the least squares regression, $Y=\alpha X+\mathbf{T}(X,Y)\beta_{\mathbf{T}(X,Y)}+\mathbf{0}\mathbf{S}+ \epsilon$ where $\mathbb{E}(\epsilon(X,\mathbf{Z}))=\mathbf{0}$, $\mathbf{Z}=\mathbf{T}(X,Y)\cup \mathbf{S}$ Then we can also write 
$Y=\alpha X+\mathbf{T}(X,Y)\beta_{\mathbf{T}(X,Y)}+\epsilon$ where
$\mathbb{E}(\epsilon(X,\mathbf{T}(X,Y)))=\mathbf{0}$. Hence $\mathbb{E}(\hat{\mathbf{\beta}}_{\mathbf{y}x.\mathbf{T}}) = \mathbb{E}(\hat{\mathbf{\beta}}_{\mathbf{y}x.\mathbf{Z}})=\tau_{yx}$ \qedsymbol

\subsection{Proof of Proposition 3}
\textit{Proof} As Definition 3.12 in \cite{henckel2019graphical} states, $\mathbf{O}(X,Y)$

$=\mathbf{PA}(\mathcal{G},\mathbf{CN}(\mathcal{G},X,Y))$ $ \setminus (X \cup \mathbf{DE}(\mathcal{G},\mathbf{CN}(\mathcal{G},X,Y))) $ 

$=\mathbf{PA}(Y)\setminus X$   \qedsymbol

\subsection{Proof of Proposition 4}
\textit{Proof}
Since the $\mathbf{Z},X,Y$ follows a structural equation model depicted in Section 2.3, we know that $Y=\beta\mathbf{PA}(\mathcal{G},Y)+\epsilon$ where $\epsilon \independent \mathbf{PA}(\mathcal{G},Y)$. Along with the uniqueness of the least squares regression, $\mathbf{PA}(\mathcal{G},Y)$ is the unique solution to (1). \qedsymbol

\subsection{Proof of Proposition 5}
\textit{Proof} 
Since $Y \independent \mathbf{Z} \setminus \mathbf{Z}'|X\cup \mathbf{Z}'$, $\mathbf{Z}'$ is a valid adjustment set relative to $X,Y$ by Theorem 3.4.1 of \cite{pearl2009causality}.
By Theorem 3.4 of \cite{henckel2019graphical}, two valid adjustment sets $\mathbf{Z}'=\mathbf{S} \cup \mathbf{T}(X,Y)$ and $\mathbf{Z}$ have that $Y \independent \mathbf{Z} \setminus \mathbf{Z}'|X\cup \mathbf{T}(X,Y)$, which is given by definition of $\mathbf{Z}'$ as well as $X \independent \emptyset |\mathbf{Z}$, then the asymptotic variance of causal effect estimators satisfy :$\textbf{var}(\hat{\mathbf{\beta}}_{\mathbf{y}x.\mathbf{Z}}) \ge \textbf{var}(\hat{\mathbf{\beta}}_{\mathbf{y}x.\mathbf{T}})$ \qedsymbol

\subsection{Proof of Theorem 1}

Since we aim to tackle the optimisation problem with objective being a non-negative convex function on constricted support, we hereby adapt the same denotation displayed in \cite{bertsimas2016best}, where we call the convex function 
$\min ||Y-\beta (X,\mathbf{Z}^T)^T||_2^2$ as $g(\beta)$. By Proposition 6(a) of \cite{bertsimas2016best}, under the constraint on $||\beta||_0 \leq k$ of a given fixed constant $k$, MIO algorithm\cite{bertsimas2016best} would produce sequence of estimators $\beta_m,m=\{1,2,3,...\}$ in its procedure, which satisfies the following property: 
$g(\beta_{m})$ decreases and converges with $g(\beta_m)-g(\beta_{m+1}) \geq \frac{L-l}{2} ||\beta_m-\beta_{m+1}||_2^2$, where $L$ is a parameter in MIO algorithm while $l=\lambda_{max}(X,\mathbf{Z})^T(X,\mathbf{Z})$, the largest eigenvalue of $(X,\mathbf{Z})^T(X,\mathbf{Z})$. Therefore, for any $L \geq l$, we will get $\beta_m$ converges. Furthermore, Proposition 6(c) guarantee that after finite iterations, MIO algorithm would arrive at a converged $\beta_m$. Therefore, we have shown that through the MIO algorithm, the unique solution to the optimizing problem $\min ||Y-\beta (X,\mathbf{Z}^T)^T||_2^2$ s.t. $||\beta||_0 \leq k$ is obtained. 

The remainder of the Algorithm 1 aims to resolve the causal constraints. Suppose that the output of Algorithm 1 is $\mathbf{Z}'$. Then the cardinality of $\mathbf{Z}'$ can't be smaller than that of $\mathbf{Pa}(Y)$. Otherwise, $\exists Z_1 \in \mathbf{Pa}(Y), \notin \mathbf{Z}'$ and $Z_1 \notindependent Y |A, \forall A\subseteq \mathbf{Z}^{m-1}$. Clearly, $g(\beta_m,Z_1) \leq g(\beta_{m+1})$, contradicting with optimality of MIO algorithm. Hence, $\mathbf{Z}'$ must contain all of $\mathbf{Pa}(Y)$. But in the procedure of Algorithm 1, any $Z \notin \mathbf{Pa}(Y)$ has $Z \independent Y |\mathbf{Pa}(Y)$, which means is pruned out already. Hence, we have established that $\mathbf{Z}'=\mathbf{Pa}(Y)$

\bibliographystyle{unsrt}  
\bibliography{template}

\begin{thebibliography}{10}

\bibitem{guo2022efficient}
F~Richard Guo and Emilija Perkovi{\'c}.
\newblock Efficient least squares for estimating total effects under linearity
  and causal sufficiency.
\newblock {\em The Journal of Machine Learning Research}, 23(1):4503--4543,
  2022.

\bibitem{henckel2019graphical}
Leonard Henckel, Emilija Perkovi{\'c}, and Marloes~H Maathuis.
\newblock Graphical criteria for efficient total effect estimation via
  adjustment in causal linear models.
\newblock {\em arXiv preprint arXiv:1907.02435}, 2019.

\bibitem{rotnitzky2019efficient}
Andrea Rotnitzky and Ezequiel Smucler.
\newblock Efficient adjustment sets for population average treatment effect
  estimation in non-parametric causal graphical models.
\newblock {\em arXiv preprint arXiv:1912.00306}, 2019.

\bibitem{witte2019covariate}
Janine Witte and Vanessa Didelez.
\newblock Covariate selection strategies for causal inference: Classification
  and comparison.
\newblock {\em Biometrical Journal}, 61(5):1270--1289, 2019.

\bibitem{wang2012bayesian}
Chi Wang, Giovanni Parmigiani, and Francesca Dominici.
\newblock Bayesian effect estimation accounting for adjustment uncertainty.
\newblock {\em Biometrics}, 68(3):661--671, 2012.

\bibitem{shortreed2017outcome}
Susan~M Shortreed and Ashkan Ertefaie.
\newblock Outcome-adaptive lasso: variable selection for causal inference.
\newblock {\em Biometrics}, 73(4):1111--1122, 2017.

\bibitem{talbot2015bayesian}
Denis Talbot, Genevieve Lefebvre, and Juli Atherton.
\newblock The bayesian causal effect estimation algorithm.
\newblock {\em Journal of Causal Inference}, 3(2):207--236, 2015.

\bibitem{pearl2009causality}
Judea Pearl.
\newblock {\em Causality}.
\newblock Cambridge university press, 2009.

\bibitem{lauritzen1996graphical}
Steffen~L Lauritzen.
\newblock {\em Graphical models}, volume~17.
\newblock Clarendon Press, 1996.

\bibitem{spirtes2000causation}
Peter Spirtes, Clark~N Glymour, Richard Scheines, and David Heckerman.
\newblock {\em Causation, prediction, and search}.
\newblock MIT press, 2000.

\bibitem{bollen2014structural}
Kenneth~A Bollen.
\newblock {\em Structural equations with latent variables}, volume 210.
\newblock John Wiley \& Sons, 2014.

\bibitem{de2011covariate}
Xavier De~Luna, Ingeborg Waernbaum, and Thomas~S Richardson.
\newblock Covariate selection for the nonparametric estimation of an average
  treatment effect.
\newblock {\em Biometrika}, 98(4):861--875, 2011.

\bibitem{glymour2008methodological}
M~Maria Glymour, Jennifer Weuve, and Jarvis~T Chen.
\newblock Methodological challenges in causal research on racial and ethnic
  patterns of cognitive trajectories: measurement, selection, and bias.
\newblock {\em Neuropsychology review}, 18(3):194--213, 2008.

\bibitem{textor2011dagitty}
Johannes Textor, Juliane Hardt, and Sven Kn{\"u}ppel.
\newblock Dagitty: a graphical tool for analyzing causal diagrams.
\newblock {\em Epidemiology}, 22(5):745, 2011.

\bibitem{li2005model}
Lexin Li, R~Dennis~Cook, and Christopher~J Nachtsheim.
\newblock Model-free variable selection.
\newblock {\em Journal of the Royal Statistical Society: Series B (Statistical
  Methodology)}, 67(2):285--299, 2005.

\bibitem{bertsimas2016best}
Dimitris Bertsimas, Angela King, and Rahul Mazumder.
\newblock Best subset selection via a modern optimization lens.
\newblock {\em The annals of statistics}, 44(2):813--852, 2016.

\bibitem{ertefaie2018variable}
Ashkan Ertefaie, Masoud Asgharian, and David~A Stephens.
\newblock Variable selection in causal inference using a simultaneous
  penalization method.
\newblock {\em Journal of Causal Inference}, 6(1), 2018.

\end{thebibliography}

\end{document}